\documentclass[12pt,epsfig,wrapfig,rotating]{article}
\usepackage{epsfig}

\newcommand{\mb}{M_{\rm bc}}
\clearpage
\begin{document}

\vspace*{-2cm} 
\hspace*{11cm} {\bf UCHEP-04-03}

\vspace*{1cm}

\begin{center}
{\Large \bf 
$\phi_2$ Measurement Using $B \to \rho\pi$ decays at Belle }

\vspace*{1cm}
{A.~Somov  \small{(for the Belle Collaboration)}} 

\vspace*{0.5cm}

{\small University of Cincinnati,Cincinnati, Ohio 45221, USA
\\E-mail: somov@bmail.kek.jp}

\end{center}
\begin{abstract}
{\small We present measurements of 
time-dependent asymmetries in decays of neutral B mesons to the final states 
$B^0 \to \rho^\pm\pi^\mp$. Measurements are based on a 140 $fb^{-1}$ data 
sample collected at the $\Upsilon(4S)$ resonance with the Belle detector at 
the KEKB asymmetric-energy $e^+e^-$ collider.}
\end{abstract}
\vspace*{0.5cm}

\section{Introduction}

The angle $\phi_2$ can be determined using a time dependent analysis in 
charmless decays such as $B^0 \to \pi^+\pi^-$, $B^0 \to \rho^+\pi^-$ and 
$B^0 \to \rho^+\rho^-$. The extraction of $\phi_2$ has been studied in 
several theoretical approaches.\cite{th_mod,isospin,Gronau} 
An isospin analysis\cite{isospin} that requires knowledge of the
branching fractions of $B^0 \to h^0h^0 (h = \pi, \rho)$ can determine
$\phi_2$ with relatively little theoretical uncertainty.

Here we present a time dependent analysis of $CP$ violation in
$B^0 \to \rho^\pm\pi^\mp$ decays. We also report on the first evidence for 
$B^0 \to \rho^0\pi^0$ decay.

\section{$B^0 \to \rho^+\pi^-$}\label{sec:rhopi}

As the final state $B^0 \to \rho^+\pi^-$ is not a $CP$ eigenstate, one has 
to consider four decay modes with different charge and flavor combinations:
$B^0 \to \rho^\pm \pi^\mp$ and $\overline{B}{}^0 \to \rho^\pm \pi^\mp$.
The decay rates can be written as
\begin{eqnarray}
& & \mathcal{P}^{\rho^\pm\pi^\mp}(\Delta t)  =  (1\pm \mathcal{A}^{\rho \pi}_{CP}) 
\frac{e^{-|\Delta t|/\tau_{B^0}}}{8\tau_{B^0}} \nonumber \\ 
& & \times \{1 - q(C_{\rho\pi}\pm\Delta C_{\rho\pi})\cos(\Delta m_d 
\Delta t)\nonumber \\
& & + q(S_{\rho\pi}\pm\Delta S_{\rho\pi})\sin(\Delta m_d \Delta t)\},
\label{eq:rate}
\end{eqnarray}
where $\Delta t = t_{\rho\pi} - t_{tag}$ is the proper-time difference
between the fully reconstructed and the associated $B$ decay, 
and $q = +1(-1)$ corresponds to 
$B^0 (\overline{B}{}^0)$ tags. The parameters $S_{\rho\pi}$ and $C_{\rho\pi}$
are associated with mixing-induced $CP$ violation (related to $\phi_2$)
and flavor-dependent direct $CP$ violation, respectively. The parameters
$\Delta S_{\rho\pi}$ and $\Delta C_{\rho\pi}$ are $CP$-conserving. 
$\Delta C_{\rho\pi}$ describes the asymmetry between the rates 
$\Gamma (B^0 \to \rho^+\pi^-) + \Gamma (\overline{B}{}^0 \to \rho^-\pi^+)$
and $\Gamma (B^0 \to \rho^-\pi^+) + \Gamma (\overline{B}{}^0 \to \rho^+\pi^-)$.
$\Delta S_{\rho\pi}$ depends in addition on difference in strong phases 
between the amplitudes contributing to $B \to \rho\pi$ decays. 
The direct $CP$ violation is related to the time and flavor integrated 
charge asymmetry, 
$\mathcal{A}_{CP}$, defined as
\begin{equation}
\mathcal{A}^{\rho \pi}_{CP} = \frac{N(\rho^+ \pi^-) - N(\rho^-\pi^+)}
{N(\rho^+ \pi^-) + N(\rho^-\pi^+)},
\label{eq:acp}
\end{equation}
where $N(\rho^+\pi^-)$ and $N(\rho^-\pi^+)$ are the sum of the yields for
$B^0$ and $\overline{B}{}^0$ decays to $\rho^+\pi^-$ and $\rho^-\pi^+$, 
respectively. We also define two separate direct-$CP$ related asymmetries:
\begin{eqnarray}
\mathcal{A}^{\rho \pi}_{\pm\mp} & = &
\frac{N(\overline{B}{}^0 \to  \rho^\mp \pi^\pm) - N(B^0 \to \rho^\pm \pi^\mp)}
{N(\overline{B}{}^0 \to  \rho^\mp \pi^\pm) + N(B^0 \to \rho^\pm \pi^\mp)} 
\nonumber \\
& = & \mp \frac{\mathcal{A}^{\rho \pi}_{CP} \pm C_{\rho\pi} \pm 
\mathcal{A}^{\rho \pi}_{CP}\Delta C_{\rho\pi}}
{1\pm \Delta C_{\rho\pi} \pm \mathcal{A}^{\rho \pi}_{CP} C_{\rho\pi}}.
\label{eq:direct}
\end{eqnarray}

The analysis is based on the quasi-two-body approach. We consider
$B^0 \to (\pi^\pm\pi^0)\pi^\mp$ candidates from two distinct bands in 
the $\pi^+\pi^-\pi^0$ Dalitz plot. The $B^0\to \rho^\pm\pi^\mp$ candidates 
are selected by requiring two oppositely charged pion-candidate tracks  
with $\pi^0$ candidates.  Charged tracks are required to originate from 
the interaction point and have transverse momenta greater than 100 MeV/c. 
To identify tracks as charged pions, we combine information from 
time-of-flight counters,  aerogel threshold cherenkov counters and the central 
tracking chambers. Neutral pion candidates are reconstructed from photon 
pairs with invariant masses in the range 
$0.118{\,{\rm GeV}/c^2}<M_{\gamma\gamma}<0.150{\,{\rm GeV}/c^2}$, and
momentum larger than 200 MeV/$c$ in the laboratory frame.
Photon candidates are selected with a minimum energy requirement of 50 MeV 
in the barrel region of the electromagnetic calorimeter 
$(32^\circ < \theta_{\gamma} < 129^\circ)$ and 100 MeV in the endcap regions, 
where $\theta_{\gamma}$ denotes the polar angle of the photon with respect 
to the beam line. We also apply a cut on the angle between the photon 
flight direction and the boost direction from the laboratory system in 
the $\pi^0$ rest frame, $|\cos \theta_{dec}^{\pi^0}| < 0.95$. 

\begin{figure}
\begin{center}
\epsfxsize190pt
\resizebox{!}{0.25\textwidth}{\includegraphics{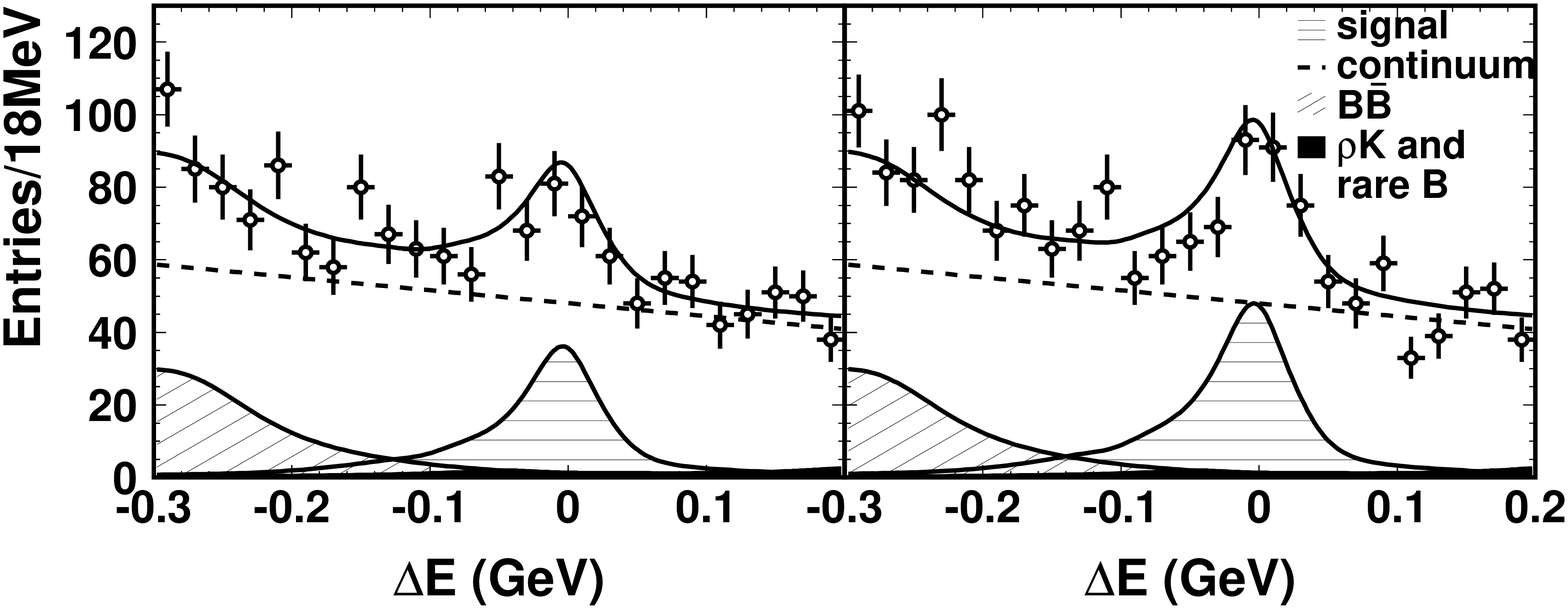}}
\resizebox{!}{0.25\textwidth}{\includegraphics{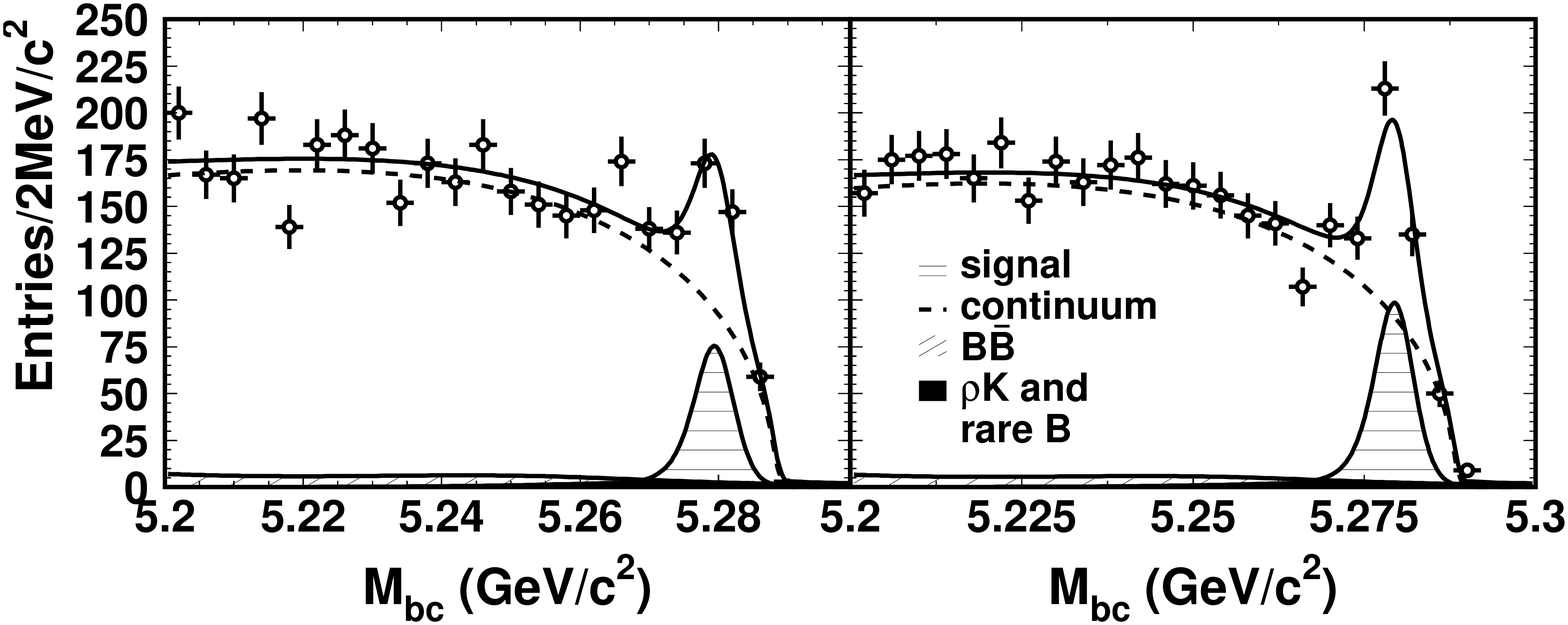}}
\caption{Projections for the results of the 2-D unbinned likelihood fit to 
the $\Delta E (top)$ and $\mb$ (bottom) distributions for $\rho^+\pi^-$  
candidates (left) and $\rho^-\pi^+$ candidates (right).}
\label{fig:mbc_de}
\end{center}
\end{figure}

$B$ meson candidates are identified using two kinematic variables:
the beam energy constrained mass $M_{bc}\equiv \sqrt{E_{beam}^2 - p_B^2}$,
and the energy difference $\Delta E \equiv E_B - E_{beam}$, where
$E_{beam}$ is the center-of-mass (CM) beam energy, and $E_B$ and $p_B$ are
the reconstructed energy and momentum of the $B$ candidate in the CM frame.
We accept events in the region $\mb > 5.21{\,{\rm GeV}/c^2}$
and  $-0.3{\,{\rm GeV}} < \Delta E < 0.2{\,{\rm GeV}}$
and define a signal region as  $\mb > 5.27{\,{\rm GeV}/c^2}$
and $-0.1{\,{\rm GeV}} < \Delta E < 0.08{\,{\rm GeV}}$.
To avoid the region where the $\rho^+\pi^-$ and $\rho^-\pi^+$ amplitudes
interfere, we exclude candidates with both $M_{\pi^\pm\pi^0}$ smaller than
1.22 GeV/$c^2$.

The dominant background to $B^0 \to \rho^+\pi^-$ is $e^+e^- \to q\overline q $ 
$(q = u,d,s,c)$  continuum events. To discriminate the signal from the 
background we use the event topology, which tends to be isotropic for 
$B\overline B$ events and jet-like for $q \overline q$ events. We combine a 
Fisher discriminant based on five modified Fox-Wolfram moments\cite{fox} 
with the cosine of the angle between the flight direction of the $B$ in the CM 
frame and the electron beam direction to form the likelihood ratio 
$\mathcal{R}=\mathcal{L}_{B\overline B}/\mathcal{L}_{B\overline B}+
\mathcal{L}_{q\overline q}$. Here, $\mathcal{L}_{B\overline B}$ and  
$\mathcal{L}_{q\overline q}$ are likelihood functions for signal and continuum
 events, respectively. The continuum background is reduced by applying a 
cut $\mathcal{R} > 0.8$.

The $b$-flavor of the accompanying $B$ meson is identified using tracks 
not associated with the reconstructed $B^0 \to \rho^\pm\pi^\mp$ decays.
The tagging algorithm is identical to that used for the Belle sin$2\phi_1$ 
measurement\cite{Kakuno:2004cf}. The tagging information is represented by two 
parameters, $q =\pm1$, the flavor, and a quality factor $r$ that ranges 
from $r=0$ for no flavor discriminant to $r=1$ for unambiguous 
flavor assignment.   

The signal-side tracks and the tag-side tracks are fit for the signal  
($z_{CP}$) and tag ($z_{tag}$) side  vertices using an interaction point 
constraint. Taking into account that $B \overline B$ are produced practically 
at rest in the $\Upsilon(4S)$ CM frame, the proper time difference 
$\Delta t$ is proportional to the distance between vertices 
$\Delta z = z_{cp} - z_{tag}$: $\Delta t \sim \Delta z/\beta \gamma c$, 
where $\beta\gamma = 0.425$ is the Lorentz boost of the $e^+e^-$ system.  

\begin{figure}
\begin{center}
\epsfxsize190pt
\resizebox{!}{0.35\textwidth}{\includegraphics{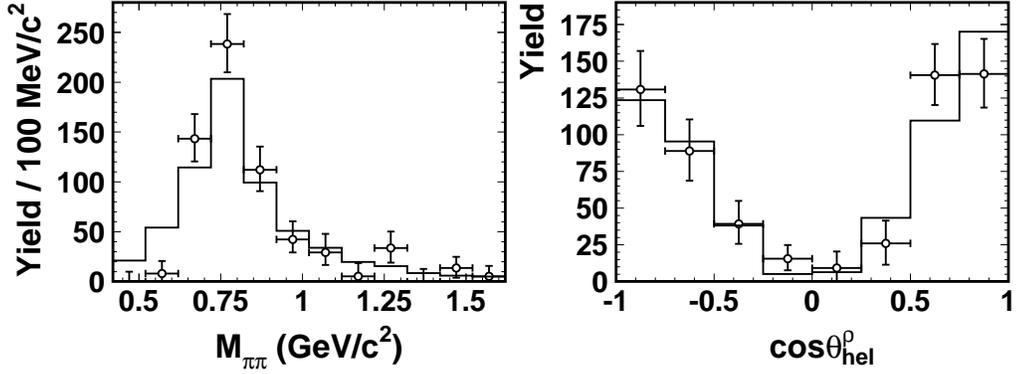}}
\caption{Signal yields as functions of $M_{\pi^\pm\pi^0}$ (left) and
$\cos\theta_{hel}^\rho$ (right). Histograms are expectations
from the $B \to \rho^\pm\pi^\mp$ MC simulation.}
\label{fig:yields}
\end{center}
\end{figure}

The $\rho^\pm\pi^\mp$ yields are extracted using an unbinned 
maximum-likelihood (ML) fit to the two-dimensional ($\mb,\Delta E$) 
distribution. Four types of background are considered: continuum 
$q\overline q$, charm $B$ background ($b \to c$), $B\to \rho K$, and rare, 
charmless $B$ decays (rare $B$). The $\mb$ and $\Delta E$ distributions in the 
signal region along with projections of the unbinned ML fit  are shown 
in Fig.~\ref{fig:mbc_de}. The fit yields $483\pm46$ $B^0 \to \rho^\pm\pi^\mp$ 
candidates and a time and flavor integrated charge asymmetry 
$\mathcal{A}^{\rho\pi}_{CP} = -0.16\pm 0.10(stat)$. Fig.~\ref{fig:yields} 
shows signal yields extracted from the fits to the ($\mb,\Delta E$) 
distributions for different $M(\pi^\pm\pi^\mp)$ and $\cos \theta_{hel}^\rho$ 
mass and helicity bins respectively. The result of $B \to \rho^\pm\pi^\mp$ MC 
simulation are shown as histograms.

The $CP$ parameters $C$, $\Delta C$, $S$, $\Delta S$ are obtained from an
unbinned ML fit to the $\Delta t$ distribution. The likelihood function
is defined as
\begin{eqnarray}
\mathcal{L} & = & \prod_{i=1}^{N}\{ f_{\rho\pi}P_{\rho\pi}(\Delta t_i) 
+ f_{\rho\pi}^{WC}P_{\rho\pi}^{WC}(\Delta t_i) \nonumber \\
& + & f_{\rho K}P_{\rho K}(\Delta t_i) + f_{q\bar q}P_{q\bar q}(\Delta t_i) 
\nonumber \\
& + & f_{B \overline B}P_{B \overline B}(\Delta t_i) 
+ f_{rare}P_{rare}(\Delta t_i)\},
\label{eq:likelihood}
\end{eqnarray}
where $f_{\rho\pi,\rho K,q\bar q,B \overline B,rare}$ is the fraction of
events determined on an event-by-event basis as a function of the
$\mb$ and $\Delta E$ for each flavor tagging $r$ interval  and $\pi^0$ 
momentum range. $P_{\rho\pi}(\Delta t_i)$ and $P_{\rho\pi}^{WC}(\Delta t_i)$
are the probability density functions (PDF) for signal reconstructed
with right-sign $\rho$ charge and wrong-sign $\rho$ charge. 
They are determined  as the convolution of the true PDF defined 
in Eq.~(\ref{eq:rate}) with the $\Delta t$  resolution function.\footnote{
the true PDF given by Eq.(\ref{eq:rate}) is modified to take into account 
the effect of incorrect flavor tagging} For $P_{\rho K}(\Delta t)$ we assume
$C = S = \Delta S = 0$, $\Delta C = -1$, and $\mathcal{A}^{\rho K}_{CP} = 0$. 
The PDF for continuum background events is 
$P_{q\bar q}(\Delta t) = \{(f_\tau/2\tau_{bkg})e^{-|\Delta \tau|/
\tau_{bkg}} + (1 - f_\tau)\delta(\Delta t)\}/2$ convoluted with a resolution 
function $R$, where $f_\tau$ is the fraction of background with the effective 
lifetime $\tau_{bkg}$, and $\delta$ is the Dirac delta function. The 
resolution function $R$ for the background is parameterized as the sum of two 
Gaussian distributions. The parameters for the PDFs are determined from the 
sideband data for continuum and from the MC simulation for the $B\overline B$ 
and rare $B$ decays. 
\begin{figure}
\begin{center}
\epsfxsize200pt
\resizebox{!}{0.57\textwidth}{\includegraphics{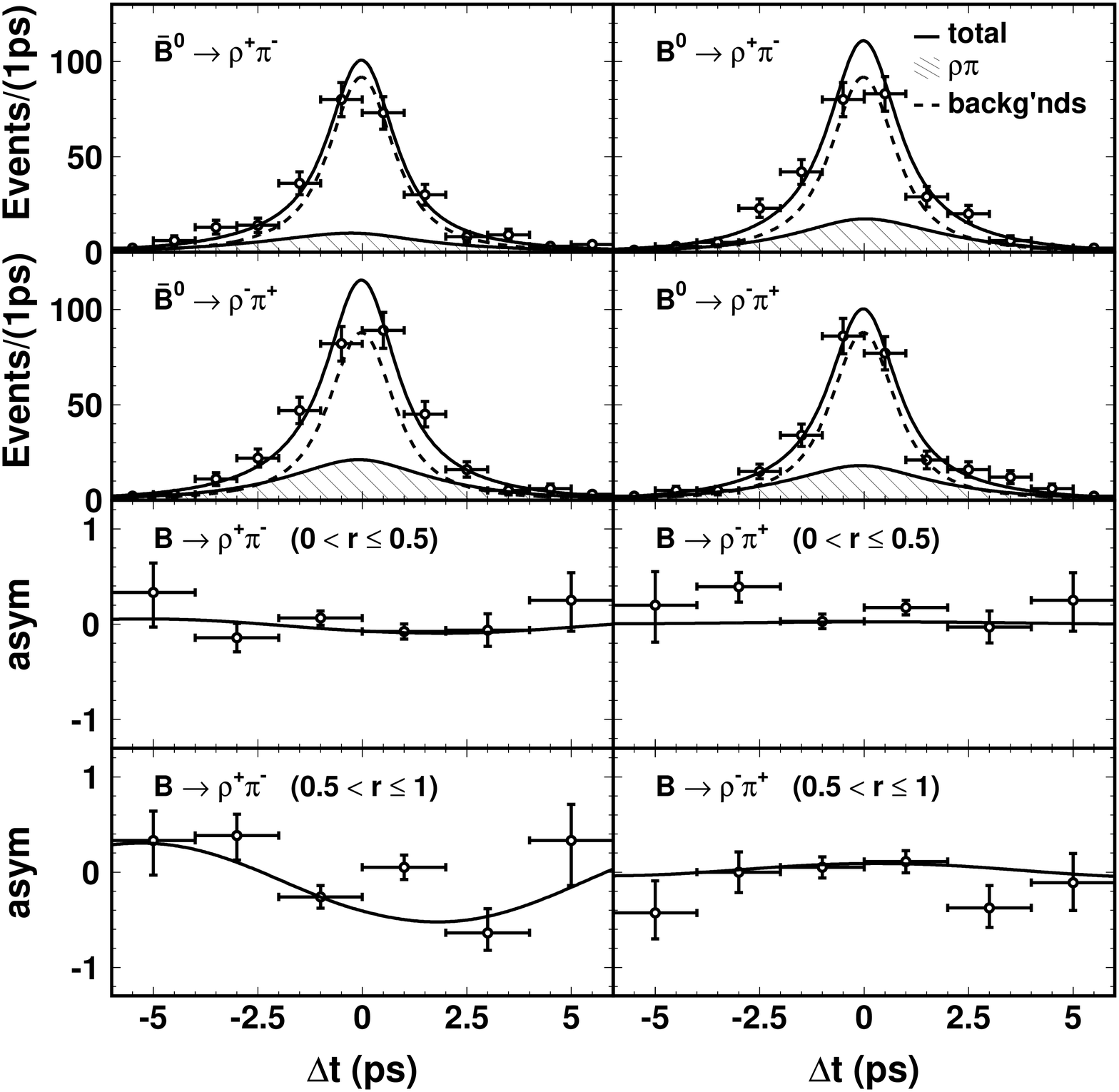}}
\caption{$\Delta t$ distributions for $B^0 \to \rho^+\pi^-$ (top) and
$B^0 \to \rho^-\pi^+$ (second row), and resulting $CP$ asymmetry (bottom rows).
The $CP$ asymmetry is shown separately for high and low quality
flavor tags. The smooth curves are fit results.}
\label{fig:cpfit}
\end{center}
\end{figure}

A fit to the 1215 candidates in the signal region
(328.7 signal, 
 11.2 $B \to \rho K$, 
833  $q \bar q$, 23.3 $B \overline B$ and 18.8 rare 
$B$ events) yields $C_{\rho\pi} = 0.25\pm 0.17^{+0.02}_{-0.06}$, 
$\Delta C_{\rho\pi} = 0.38\pm 0.18^{+0.02}_{-0.04}$, 
$S_{\rho\pi} = -0.28\pm 0.23^{+0.10}_{-0.08}$, and 
$\Delta S_{\rho\pi} = -0.30\pm 0.24 \pm 0.09$. 
The first errors are statistical and the second systematic. The 
resulting $\Delta t$ distributions 
are shown in Fig.~\ref{fig:cpfit} along with projections of the fit.

The systematic errors include the uncertainties in the background fraction, 
background $\Delta t$ PDF, wrong-tag fractions, vertex reconstruction, 
$\rho\pi$ and $\rho K$ $\Delta t$ resolution functions, and fitting bias.

\section{$B^0 \to \rho^0\pi^0$}
The main reconstruction features of the analysis are the same as those for
the $B^0 \to \rho^\pm\pi^\mp$ decay channel. Neutral pion candidates are 
selected from the range $0.115{\,{\rm GeV}/c^2}<M_{\gamma\gamma}<
0.154{\,{\rm GeV}/c^2}$.
Possible contributions from ($b \to c$) background to the $\pi^+\pi^-\pi^0$ 
final state are vetoed for the decays $B^0 \to D^-\pi^+, \bar D^0\pi^0$ and
$J/\psi \pi^0$. We require the $\pi^+\pi^-$ invariant mass to be in the
range  $0.5{\,{\rm GeV/c^2}} < m_{\pi^+\pi^-} < 1.1{\,{\rm GeV/c^2}}$
and the helicity 
angle for $\rho^0$ candidates $|\cos\theta_{hel}^{\rho}| > 0.5$, where
$\theta_{hel}^{\rho}$ is defined as the angle between the $\pi^-$ direction
and the direction opposite the $B$ in the $\rho$ rest frame. 
The contributions from the $B^0 \to \rho^\pm\pi^\mp$ events are vetoed by
rejecting candidates with $\pi^\pm\pi^0$ invariant mass  
$0.5{\,{\rm GeV/c^2}} < m_{\pi^\pm\pi^0}  < 1.1{\,{\rm GeV/c^2}}$.

To reduce the dominant continuum background we use the likelihood
ratio $\mathcal{R}$ described in Sec.~\ref{sec:rhopi}.
Additional background discrimination is provided by the $b$-flavor
tagging algorithm. We utilize the tagging parameter $r$, which ranges from 0
to 1 and is a measure of the confidence that the $b$ flavor of the
accompanying $B$ meson is correctly assigned. Events with a high value of
$r$ are well-tagged and are less likely to have originated from the
continuum production.
\begin{figure}
\begin{center}
\epsfxsize190pt
\resizebox{!}{0.35\textwidth}{\includegraphics{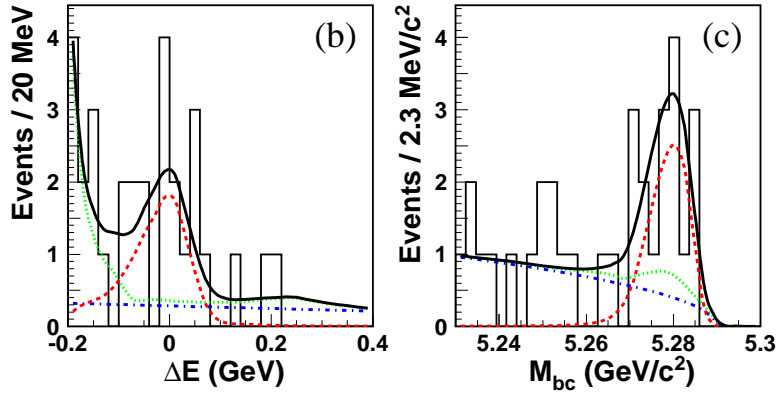}}
\caption{Distribution of $\Delta E (\mb)$ in the signal region of
$\mb (\Delta E)$. Projections of the unbinned ML fit are shown as the 
smooth curves; the dashed line represents the signal component;
the dot-dashed line is the contribution from continuum events, and 
the dotted line is the  composite of continuum and $B$-related backgrounds.}
\label{fig:mbc_de_pi0pi0}
\end{center}
\end{figure}

$B$ candidates are accepted in the region
$5.23{\,{\rm GeV/c^2}} < \mb < 5.3{\,{\rm GeV/c^2}}$,
$-0.2{\,{\rm GeV}} < \Delta E < 0.4{\,{\rm GeV}}$. The signal region is 
defined as $5.269{\,{\rm GeV/c^2}} < \mb < 5.29{\,{\rm GeV/c^2}}$
and $-0.135{\,{\rm GeV}} < \Delta E < 0.082{\,{\rm GeV}}$.
The signal yield is obtained using an unbinned ML fit to the $\mb$-$\Delta E$ 
distribution. The fit includes components for signal, continuum, $b\to c$ 
decays and the charmless $B$ decays 
$B^+ \to \rho^+\rho^0$, $B^+ \to \rho^+\pi^0$ and 
$B^+ \to \pi^+\pi^0$. The fit result is presented in 
Fig.~\ref{fig:mbc_de_pi0pi0}. The signal yield is $15\pm4.8$ events, 
corresponding to a  $3.6\sigma$ significance.

We consider systematic uncertainties from the following sources:
$\pi^0$ reconstruction efficiency ($\pm3.5\%$), track finding efficiency
($\pm2.4\%$), not included charmless $B$ decays ($\pm5.3\%$), PDF shapes 
($\pm1.6\%$), and possible interference with $B^0 \to \rho^\pm\pi^\mp$ decays
 ($9.3\%$). We obtain the branching fraction
\begin{eqnarray}
\mathcal{B}(B^0 \to \rho^0\pi^0) & = & (5.1\pm  1.6(stat) \nonumber \\
                                 &   & \pm0.9(syst))\times 10^{-6}.
\label{eq:it}
\end{eqnarray}
\section{Summary}
We have measured $CP$ violating parameters for $B^0 \to \rho^\pm\pi^\mp$ 
decays and have also obtained the first evidence for the decay
$B^0 \to \rho^0\pi^0$. The branching fraction measured is relatively
large $(5.1\pm 1.6\pm0.9)\times 10^{-6}$, and thus
the isospin analysis does not significantly limit the penguin diagram 
contribution.\cite{Stark:2003nq} A recent theoretical
model\cite{Gronau}  which assumes broken flavor-$SU(3)$ symmetry gives
$\phi_2 = (102\pm 11\pm 15)^\circ$.
The first error is experimental while the second is theoretical uncertainty.

\end{document}